\documentstyle[prd,aps,epsfig]{revtex}
\tighten
\begin{document}
\draft
%%%--uncomment for twocolumn
\twocolumn[\hsize\textwidth\columnwidth\hsize\csname
@twocolumnfalse\endcsname
%%%--
\title{COULD ELECTROMAGNETIC CORRECTIONS \\
       SOLVE THE VORTON EXCESS PROBLEM ?}
\author{Alejandro GANGUI, Patrick PETER, and C\'eline BOEHM}
\address{D\'epartement d'Astrophysique Relativiste et de Cosmologie,\\
Observatoire de Paris--Meudon, UPR 176, CNRS, 92195 Meudon, France \\
%\sevrm 
{\rm Email: 
gangui@mirabelle.obspm.fr, peter@prunelle.obspm.fr, boehm@prunelle.obspm.fr}}
\date{\today}
\maketitle
\begin{abstract}
The modifications of circular cosmic string loop dynamics due to the
electromagnetic self--interaction are calculated and shown to reduce
the available phase space for reaching classical vorton states,
thereby decreasing their remnant abundance.  
Use is made of the duality between master--function and Lagrangian 
formalisms on an explicit model. 
\end{abstract}

\vskip2pc
%%%--uncomment for twocolumn
]
%%%--

\section{Introduction}

Most particle physics theories, extensions of the so--called standard
model of interactions, suggest that topological defects~\cite{kibble}
should have been formed during phase transitions in the early
universe~\cite{book}. Among those, the most fashionable, because of
their ability to solve many cosmological puzzles, are cosmic strings,
provided couplings with other particles are such that they are not of
the superconducting kind~\cite{witten}. In the latter case, however,
they would not be able to decay into pure gravitational radiation,
terminating their life in the form of frozen vorton states which might
be so numerous that they would cause a cosmological
catastrophe~\cite{DS,vorton,vortons}: a rough evaluation of their
abundance yields a very stringent constraint of a symmetry breaking
scale which, to avoid an excess, should be less than $10^9$ GeV, which
is incompatible with the idea of them being responsible for galaxy
formation and leaving imprints in the cosmic microwave background.

There are many ways out of this vorton excess problem, the most widely
accepted relying upon stability considerations: since vortons are
centrifugally supported string loop configurations, the origin of the
rotation being hidden in the existence of a current, it is legitimate
to first ask whether the current itself is stable against decay by
quantum tunnelling. This question, however, has not yet been properly
addressed, and presumably depends on the particular underlying field
model one uses, so that, although it is clearly an important point to
be clarified, it will not be considered in this work. Another issue,
at a lower level, concerns the classical stability: a rotating string
configuration in equilibrium may exhibit unstable perturbation states;
if it were the case in general for any equation of state, then one
would expect vortons to dissipate somehow, and the problem would be
cured~\cite{Carter-Martin,Martin}. This hope is not however fulfilled
by the Witten kind of strings whose equation of
state~\cite{neutral,enon0} falls into the possibly stable
category~\cite{Martin-Peter}.

Finally, another point worth investigating is that of vorton
formation. It should be clear that an arbitrary cosmic string loop,
endowed with fixed ``quantum'' numbers, will not in general end up in
the form of a vorton. This has to be quantified somehow, and one way
of doing so is achieved by looking at some specific initial
configuration, circular say, and then letting it evolve until it
reaches an equilibrium state, if any. This has been
done~\cite{Larsen-Axenides,US} for various neutral current--carrying
cases which showed again that many loops can indeed end up as vorton
states, and using many different equations of
state~\cite{Carter-Peter}, moreover providing analytic solutions to
the elastic kind of string equations which may be useful in future
sophisticated numerical simulations taking into account the possible
existence of currents.

Our purpose here is twofold. First, it is our aim to calculate the
effect of including an electromagnetic self--coupling in the dynamics
of a rotating loop. The reason for doing it is that this
self--interaction modifies the equation of
state~\cite{enon0,emCarter}, so that the evolution is indeed supposed
to be different. Besides, it is the very first correction that can be
included without the bother of introducing the much more troublesome
complications of evaluating radiation, and in fact the radiation can
only be consistently calculated provided this first order effect has
been properly taken into account.

The development of the formalism needed to make these calculations is
also among the motivations for this work: the usual way of working out
the dynamics of a current--carrying string (or any worldsheet of
arbitrary dimension living in a higher dimensional space) consists in
varying an action which is essentially the integral over the
worldsheet of a Lagrangian function, itself seen as a function of the
squared gradient of a scalar function living on the worldsheet and
representing the variations of the actual phase of some physical
field~\cite{witten,mal}. There is however an ambiguity in this
procedure in the sense that the phase gradient used can be chosen in a
different way by means of a Legendre kind of
transformation~\cite{Formalism} whereby one then considers the
relevant dynamical variable to be instead the current itself. This
newer alternative procedure provides a completely equivalent dual
formulation which turns out to be the only one that can deal with some
instances like that of inclusion of the electromagnetic corrections
here considered.

In the following section, we recapitulate this duality between both
descriptions, whose equivalence we show explicitely. Then,
after a brief description of how are electromagnetic self--corrections
included, we discuss the particular case of a circular rotating loop
for which we calculate an effective potential in view of resolving the
dynamics. We conclude by showing that in general this corrective
effect tends to reduce the number of vorton states attainable for
arbitrary initial conditions.

\section{The dual formalism}

The usual procedure for treating a specific cosmic string dynamical
problem consists in writing and varying an action which is assumed to
be the integral over the worldsheet of a Lagrangian function depending
on the internal degrees of freedom of the worldsheet. In particular,
for the structureless string, this is taken to be the Goto--Nambu
action~\cite{GN}, i.e. the integral over the surface of the constant
string tension. In more general cases, various functions have been
suggested that supposedly apply to various microscopic field
configurations~\cite{Carter-Peter}. They share the feature that the
description is achieved by means of a scalar function $\varphi$,
identified with the phase of a physical field trapped on the string,
whose squared gradient, called the state parameter $w\propto\partial
_a \varphi \partial ^a\varphi$ ($a$ denoting a string coordinate
index), has values which completely determine the dynamics through a
Lagrangian function ${\cal L}\{ w\}$. This description has the
pleasant feature that it is easily understandable, given the clear
physical meaning of $\varphi$. However, as we shall see, there are
instances for which it is not so easily implemented and for which an
alternative, equally valid, formalism is better
adapted~\cite{Formalism}.

In this section we will derive in parallel expressions for the
currents and state parameters in these two representations, which are
dual to each other.  This will not be specific to superconducting
vacuum vortex defects, but is generally valid to the wider category of
elastic string models~\cite{Formalism}.  In this formalism one works
with a two--dimensional worldsheet supported master function $\Lambda
\{ \chi \}$ considered as the dual of ${\cal L}\{ w \}$, these functions
depending respectively on the squared magnitude of the gauge
covariant derivative of the scalar potentials $\psi$ and $\varphi$ as
given by
\begin{equation}
\chi ={\tilde\kappa}_{_0}\gamma^{ab}\psi_{|a }\psi_{|b} \ \
\longleftrightarrow \ \ 
w =\kappa_{_0}\gamma^{ab}\varphi_{|a }\varphi_{|b} \ ,
\label{state-parameters}
\end{equation}
where $\kappa_{_0}$ and ${\tilde\kappa}_{_0}$ are adjustable 
respectively positive and negative dimensionless normalisation 
constants that, as we will see below, are related to each other.  
The arrow in the previous equation stands to mean an exact 
correspondence between quantities appropriate to each dual representation. 
We use the notation
$\gamma^{ab}$ for the inverse of the induced metric, $\gamma_{ab}$ on
the worldsheet. The latter will be given, in terms of the background
spacetime metric $g_{\mu\nu}$ with respect to the 4--dimensional
background coordinates $x^\mu$ of the worldsheet, by
\begin{equation}
 \gamma_{ab}=g_{\mu\nu} x^\mu_{\, ,a} x^\nu_{\, ,b} \ ,
\end{equation}
using a comma to denote simple partial differentiation with respect to
the worldsheet coordinates $\xi^a$ and using Latin indices for the
worldsheet coordinates $\xi^{_1}=\sigma$ (spacelike),
$\xi^{_0}=\tau$ (timelike).  The gauge covariant derivative
$\varphi_{|a}$ would be expressible in the presence of a background
electromagnetic field with Maxwellian gauge covector $A_\mu$ by
$\varphi_{|a}=\varphi_{,a}\! -\! eA_\mu x^\mu_{\, ,a}$.

In Eq.~(\ref{state-parameters}) the scalar potentials $\psi$ and
$\varphi$ are such that their gradients are orthogonal to each other,
namely
\begin{equation}
\gamma^{ab}\varphi_{|a }\psi_{|b} = 0 \ ,
\label{ortho}
\end{equation}
implying that if one of the gradients, say $\varphi_{|a }$ is
timelike, then the other one, say $\psi_{|a}$, will be spacelike, which
explains the different signs of the dimensionless constants
$\kappa_{_0}$ and ${\tilde\kappa}_{_0}$.

Whether or not background electromagnetic and gravitational fields are
present, the dynamics of the system can be described in the two
equivalent dual representations~\cite{Formalism,mal} which are
governed by the master function $\Lambda$ and the Lagrangian scalar
${\cal L}$, that are functions only of the state parameters $\chi$ and
$w$, respectively.  The corresponding conserved current
vectors, $n^a$ and $z^a$ say, in the worldsheet, will be given
according to the Noetherian prescription
\begin{equation}
n^a=- {\partial {\Lambda}\over\partial \psi_{|a} }\ \ 
\longleftrightarrow \ \ 
z^a=- {\partial {\cal L}\over\partial \varphi_{|a} }\ .
\end{equation}
This implies
\begin{equation}
\tilde {\cal K} n^a= {\tilde\kappa}_{_0} \psi^{|a}\ \ 
\longleftrightarrow \ \ 
{\cal K} z^a=  \kappa_{_0} \varphi^{|a}\ ,
\label{zcur-}
\end{equation}
where we use the induced metric for internal index raising, and where
${\cal K}$ and $\tilde {\cal K}$ can be written as
\begin{equation}
\tilde {\cal K}^{-1} = - 2 { d{\Lambda}\over d\chi }\ \ 
\longleftrightarrow \ \ 
{\cal K}^{-1} = -2 { d{\cal L} \over dw} .
\label{calk-}
\end{equation}
As it will turn out, the equivalence of the two mutually dual descriptions
is ensured provided the relation
\begin{equation} \tilde {\cal K} = -{\cal K}^{-1},\label{KK}
\end{equation}
holds. This means one can define ${\cal K}$ in two alternative ways,
depending on whether it is seen it as a function of $\Lambda$ or of
${\cal L}$. We shall therefore no longer use the function $\tilde
{\cal K}$ in what follows.

The currents $n^a$ and $z^a$ {\it in} the worldsheet can be
represented by the corresponding tangential current vectors $n^\mu$
and $z^\mu$ {\it on} the worldsheet, where the latter are defined with
respect to the background coordinates, $x^\mu$, by
\begin{equation}
n^\mu=n^a x^\mu_{\, ,a} \ \ 
\longleftrightarrow \ \ 
z^\mu=z^a x^\mu_{\, ,a} \ .
\label{back-ground}
\end{equation}

The purpose of introducing the dimensionless scale constants
${\tilde\kappa}_{_0}$ and $\kappa_{_0}$ is to simplify macroscopic
dynamical calculations by arranging for the variable coefficient
${\cal K}$ to tend to unity when $\chi$ and $w$ tend to zero, i.e. in
the limit for which the current is null.  To obtain the desired
simplification it is convenient not to work directly with the
fundamental current vectors $n^\mu$ and $z^\mu$ that (in units such
that the Dirac Planck constant $\hbar$ is set to unity) will represent
the quantized fluxes, but to work instead with the
corresponding rescaled currents $\omega^\mu$ and $c^\mu$ that
are obtained by setting
\begin{equation}
n^\mu=\sqrt{-{\tilde\kappa}_{_0}}\,\omega^\mu \ \ 
\longleftrightarrow \ \ 
z^\mu=\sqrt{\kappa_{_0}}\, c^\mu \ .
\label{scur-}\end{equation}

Based on Eq.~(\ref{ortho}) that expresses the orthogonality of 
the scalar potentials we can conveniently write
the relation between $\psi$ and $\varphi$ as follows
\begin{equation}
\varphi_{|a } = {\cal K}
{\sqrt{-{\tilde\kappa}_{_0}}\over \sqrt{\kappa_{_0}}} \epsilon_{ab}
\psi^{|b} \ ,
\label{ansatzzz}
\end{equation}
where $\epsilon$ is the antisymmetric surface measure tensor (whose
square is the induced metric, $\epsilon_{ab}\epsilon^b{_ c}
=\gamma_{ac}$).
From this and using Eq. (\ref{state-parameters}) we easily get 
the relation between the state variables,
\begin{equation} w={\cal K}^2\chi .\label{wK} \end{equation}
In terms of the rescaled currents, and using Eqs. (\ref{zcur-}) and
(\ref{back-ground}) we get
\begin{equation}
c_\mu c^\mu           = w / {\cal K}^2 = \chi    \ \ 
\longleftrightarrow \ \ 
\omega_\mu \omega^\mu = - {\cal K}^2  \chi = - w \ .
\label{current-vectors}
\end{equation}

Both the master function $\Lambda$ and the Lagrangian ${\cal L}$ are
related by a Legendre type transformation that gives
\begin{equation}
\Lambda={\cal L}+{\cal K}\chi\ .
\label{Lamb-} \end{equation}
The functions ${\cal L}$ and $\Lambda$ can be seen~\cite{Formalism} to
provide values for the energy per unit length $U$ and the tension $T$
of the string depending on the signs of the state parameters $\chi$
and $w$. (Originally, analytic forms~\cite{Carter-Peter} for these
functions ${\cal L}$ and $\Lambda$ were derived as best fits to the
eigenvalues of the stress--energy tensor in microscopic field
theories~\cite{neutral,enon0}). The necessary identifications are
summarized in Table 1.

\begin{table}[h]
\begin{center}
%\phantom{.}
%\vspace{1cm}
\begin{tabular}{*{5}{c}}
\multicolumn{5}{c}{}\\
\multicolumn{5}{c}{\large \bf Equations of state for both regimes}\\ 
\multicolumn{5}{c}{}\\
\hline
regime & $U$ & $T$ & $\chi$ and $w$ & current\\[0.5ex]
\hline
electric & $-\Lambda$  & $-{\cal L}$ & $< 0$ & timelike\\[0.5ex]
magnetic & $-{\cal L}$ & $-\Lambda$ & $> 0$ & spacelike\\[0.5ex]
%\hline
\end{tabular}
%\vspace{-3cm}
\caption{Values of the energy per unit length $U$ and tension $T$
depending on the timelike or spacelike character of the current,
expressed as the negative values of either $\Lambda$ or ${\cal L}$.}
%\vspace{-4cm}
\end{center}
\end{table}

This way of identifying the energy per unit length and tension with
the Lagrangian and master functions also provides the constraints on
the validity of these descriptions: the range of
variation of either $w$ or $\chi$ follows from the requirement of
local stability, which is equivalent to the demand that the squared
speeds $c_{_{\rm E}}^{\, 2}=T/U$ and $c_{_{\rm L}}^{\,2} =-dT/dU$ of
extrinsic (wiggle) and longitudinal (sound type) perturbations be
positive. This is thus characterized by the unique relation
\begin{equation}
{{\cal L} \over\Lambda}>0>{d{\cal L} \over d\Lambda}\ ,
\label{stabc-} 
\end{equation}
which should be equally valid in both the electric and magnetic
ranges.

Having defined the internal quantities, we now turn to the actual
dynamics of the worldsheet and prove explicitely the equivalence
between the two descriptions.

\section{Equivalence between ${\cal L}$ and $\Lambda$.}

The claim is that the dynamical equations for the string model can be
obtained either from the master function $\Lambda$ or from the
Lagrangian ${\cal L}$ in the usual way, by applying the variation
principle to the surface action integrals
\begin{equation}
{\cal S}_\Lambda = \int d\sigma\,d\tau\,\sqrt{-\gamma}\, \Lambda \{ \chi \}
, \label{action} \end{equation}
and
\begin{equation}
{\cal S} _{\cal L} = \int d\sigma\,d\tau\,\sqrt{-\gamma}\, {\cal L}\{w\}
,\label{action2} \end{equation} 
(where $\gamma\equiv \det \{\gamma_{ab}\}$) in which the independent
variables are either the scalar potential $\psi$ or the phase field
$\varphi$ on the worldsheet and the position of the worldsheet itself,
as specified by the functions $x^\mu\{\sigma,\tau\}$.

The simplest way to actually prove this claim is to calculate
explicitely the dynamical equations and show that they yield the same
physical motion. To do this, we shall see that Eq.~(\ref{KK}) is
crucial by establishing a relation between the dynamically conserved
currents in both formalisms.

Independently of the detailed form of the complete system, one knows
in advance, as a consequence of the local or global $U(1)$ phase
invariance group, that the corresponding Noether currents will be
conserved, namely
\begin{equation}
\big(\sqrt{-\gamma}\, n^a\big)_{,a}=0\ \ 
\longleftrightarrow \ \ 
\big(\sqrt{-\gamma}\, z^a\big)_{,a}=0\ .
\end{equation}
For a closed string loop, this implies (by Green's theorem) the
conservation of the corresponding flux integrals
\begin{equation}
N=\oint d\xi^a \epsilon_{ab} n^b\ \ 
\longleftrightarrow \ \ 
Z=\oint d\xi^a \epsilon_{ab} z^b\ ,
\label{zin}
\end{equation}
meaning that for any circuit round the loop one will obtain the same
value for the integer numbers $N$ and $Z$, respectively.  $Z$ is
interpretable as the integral value of the number of carrier particles
in the loop, so that in the charge coupled case, the total electric
charge of the loop will be $Q=Ze$.

The loop will also be characterised by a second independent integer 
number $N$ whose conservation is trivially obvious. 
Thus we have the topologically conserved numbers defined by
\begin{eqnarray}
2\pi Z=\oint d\psi & = & \oint d\xi^a \psi_{|a} = \oint d\xi^a \psi_{,a}
\nonumber \\
& \longleftrightarrow & \nonumber \\
2\pi N = \oint d\varphi & = & \oint d\xi^a \varphi_{|a} = 
\oint d\xi^a \varphi_{,a} \ ,
\label{win}
\end{eqnarray}
where it is clear that $N$, being related to the phase of a physical
microscopic field, has the meaning of what is usually referred to as
the winding number of the string loop. 
The last equalities in Eqs.~(\ref{win}) follow just from explicitly 
writing the covariant derivative $_{|a}$ and noting that the circulation 
integral multiplying $A_\mu$ vanishes. 
Note however that, although $Z$
and $N$ have a clearly defined meaning in terms of underlying
microscopic quantities, because of Eqs.~(\ref{zin}) and (\ref{win}),
the roles of the dynamically and topologically conserved integer 
numbers are interchanged depending on whether we derive our equations
from $\Lambda$ or from its dual ${\cal L}$. Moreover, those two
equations, together with Eq.~(\ref{ansatzzz}) yield
\begin{equation} 4\pi^2 \kappa_{_0} \tilde \kappa_{_0} = -1,
\end{equation}
which confirms our original assumption.

As usual, the stress momentum energy density distributions $\hat
T^{\mu\nu}_\Lambda$ and $\hat T^{\mu\nu}_{\cal L}$ on the background
spacetime are derivable from the action by varying the actions with
respect to the background metric, according to the specifications
\begin{equation}
\hat T^{\mu\nu}_\Lambda \equiv {2\over\sqrt{-g}}{\delta{\cal S}_\Lambda
\over \delta g_{\mu\nu}} \equiv  {2\over\sqrt{-g}}
{\partial(\sqrt{-g}\,\Lambda )\over\partial g_{\mu\nu}}, \label{tmunu}
\end{equation}
and
\begin{equation}
\hat T^{\mu\nu}_{\cal L} \equiv {2\over\sqrt{-g}}{\delta{\cal S}_{\cal L}
\over \delta g_{\mu\nu}} \equiv  {2\over\sqrt{-g}}
{\partial(\sqrt{-g}\,{\cal L})\over\partial g_{\mu\nu}} .
\end{equation}
This leads to expressions of the standard form
\begin{equation}
\sqrt{-g}\, \hat T^{\mu\nu}=\int d\sigma\,d\tau\,\sqrt{-\gamma}\,
\delta^{(4)} \left[x^\rho - x^\rho \{\sigma,\tau \}\right]\,
\overline T{^{\mu\nu}} \end{equation}
in which the {\it surface} stress energy momentum tensors on the
worldsheet (from which the surface energy density $U$ and the string
tension $T$ are obtainable as the negatives of its eigenvalues) can be
seen to be given~\cite{Formalism,mal} by
\begin{equation}
\overline T^{\mu\nu}_\Lambda = {\Lambda}\eta^{\mu\nu} +{\cal K}^{-1} 
\omega^\mu \omega^\nu \ \ 
\longleftrightarrow \ \ 
\overline T^{\mu\nu}_{\cal L} = {\cal L}\eta^{\mu\nu} +{\cal K}
c^\mu c^\nu \ , \label{stress} 
\end{equation}
where the (first) fundamental tensor of the worldsheet is given by
\begin{equation}
\eta^{\mu\nu}= \gamma^{ab} x^\mu_{,a} x^\nu_{,b} \ .
\end{equation}

Plugging Eqs.~(\ref{scur-}) into Eqs.~(\ref{stress}), and using
Eqs.~(\ref{KK}), (\ref{wK}) and (\ref{Lamb-}), we find that the two
stress--energy tensors coincide:
\begin{equation} \overline T^{\mu\nu}_{\cal L} = \overline T^{\mu\nu}_\Lambda
\equiv \overline T^{\mu\nu}.\end{equation}
This is indeed what we were looking for since the dynamical equations
for the case at hand, namely
\begin{equation} \eta ^\rho _\mu \nabla _\rho \overline T^{\mu\nu}
=0,\end{equation} which hold for the uncoupled case, are then strictly
equivalent whether we start with the action $S_\Lambda$ or with
$S_{\cal L}$. \rule{2mm}{2mm}

\section{Inclusion of Electromagnetic Corrections}

Implementing electromagnetic corrections~\cite{emCarter}, even at the
first order, is not an easy task as can already be seen by the much
simpler case of a charged particle for which a mass renormalization is
required even before going on calculating anything in effect related
to electromagnetic field. The same applies in the current--carrying
string case, and the required renormalization now concerns the master
function $\Lambda$. However, provided this renormalization is
adequately performed, inclusion of electromagnetic corrections, 
at first order in the coupling between the current and the 
self--generated electromagnetic field, 
then becomes a very simple matter of shifting the equation
of state, everything else being left unchanged. Let us see how this
works explicitely.

Setting $K{_{\mu\nu}}^\rho \equiv \eta^\tau_\mu \eta^\sigma_\nu
\nabla _\tau \eta^\rho_\sigma$ the second fundamental tensor of the
worldsheet~\cite{Formalism}, the equations of motion of a charge
coupled string read
\begin{equation} \overline T^{\mu\nu} K{_{\mu\nu}}^\rho = \perp
^{\rho\mu}F_{\mu\nu} j^\nu,
\label{eq-mot}\end{equation} where $\perp^{\rho\mu}$ is
the tensor of orthogonal projection ($\perp^\rho_\mu = g^\rho_\mu -
\eta^\rho_\mu$), $F_{\mu\nu}=2\nabla_{[\mu} A_{\nu ]}$ the
electromagnetic tensor and $j^\mu$ the electromagnetic current flowing
along the string, namely in our case
\begin{equation} 
j^\mu = r e z^\mu \equiv q c^\mu ,\end{equation}
with $r$ the effective charge of the current carrier in unit of the
electron charge $e$ (working here in units where $e^2 \simeq 1/137$).
The self interaction electromagnetic field on the worldsheet itself
can be evaluated~\cite{witten} and one finds
\begin{equation} A^\mu\Big|_{_{string}} = \lambda j^\mu = \lambda q c^\mu,
\label{selfA}\end{equation}
with
\begin{equation} \lambda = 2 \ln (m_\sigma \Delta ),
\label{lambda}\end{equation}
where $\Delta$ is an infrared cutoff scale to compensate for the
asymptotically logarithmic behaviour of the electromagnetic
potential~\cite{enon0} and $m_\sigma$ the ultraviolet cutoff
corresponding to the effectively finite thickness of the charge
condensate, i.e., the Compton wavelength of the current-carrier
$m_\sigma^{-1}$~\cite{neutral,enon0}. In the practical situation of a
closed loop, $\Delta$ should at most be taken as the total length of
the loop.

The contribution of the self field~(\ref{selfA}) in the equations of
motion~(\ref{eq-mot}) can be calculated using the
relations~\cite{emCarter}
\begin{equation} \perp^\rho_\mu \eta^\sigma_\mu \nabla_\sigma 
 A^\nu\Big|_{_{string}} =  A^\nu\Big|_{_{string}} K{_{\mu\nu}}^\rho,
\end{equation}
and
\begin{equation} F_{\mu\nu}\Big|_{_{string}} = 2 \eta^\sigma_{[\mu} (
\nabla_\sigma +{1\over 2} \eta^{\alpha\beta} K{_{\alpha\beta}}^\sigma
) A_{\nu ]}\Big|_{_{string}},
\end{equation}
which transform Eqs.~(\ref{eq-mot}) into
\begin{equation} K{_{\mu\nu}}^\rho \left[ \overline T^{\mu\nu} + \lambda
q^2 \left( c^\mu c^\nu - {1\over 2} \eta^{\mu\nu} \chi c_\alpha
c^\alpha \right) \right] = 0,\end{equation} which is interpretable as
a renormalization of the stress energy tensor. This equation is
recovered if, in Eq.~(\ref{tmunu}), one uses
\begin{equation} \Lambda \to 
\Lambda + {1\over 2} \lambda q^2 \chi \label{start}
\end{equation}
instead of $\Lambda$. This formula~\cite{emCarter} generalises the
action renormalisation originally obtained~\cite{CHT} in the special
case for which the unperturbed model is of simple Goto--Nambu type.

That the correction enters through a simple modification of $\Lambda
\{ \chi \}$ and not of ${\cal L} \{ w\}$ is understandable if one
remembers that $\chi$ is the amplitude of the current, so that a
perturbation in the electromagnetic field acts on the current
linearly, so that an expansion in the electromagnetic field and
current yields, to first order in $q$, $\Lambda \to \Lambda + {1\over
2}j_\mu A^\mu$, which transforms easily into Eq.~(\ref{start}).

Using this modification entitles us to consider essentially non
coupled string worldsheet dynamics at this order, an uttermost
simplification since we thus do not have to consider radiation
backreaction. Note however that the correction we are now going to
take into account is necessary prior to any evaluation of the
radiation. We therefore still have to define the circular motion but
before that, let us specify the model, i.e., the equation of state
before the corrections are included.

\section{Equation of state}

The underlying field theoretical model we wish to consider is that
originally proposed by Witten to describe the current--carrying
abilities of cosmic strings~\cite{witten}. Although it is the simplest
possible model fulfilling that purpose, it is believed to share most
of the features that would be expected from more realistic
current--carrying cosmic string models~\cite{bps}. In essence, the
microscopic properties of the string are described by means of two
complex scalar fields, the string--forming symmetry--breaking Higgs
field, and the charged--coupled (or not~\cite{neutral})
current--carrier~\cite{enon0}, whose phase gradient serves to
calculate the state parameter $w$. Once these fields are defined, it
suffices to consider a stationnary and axisymmetric configuration and
integrate the corresponding relevant stress--energy tensor components
over a cross--section of the vortex to deduce the energy per unit
length and tension of the string. Repeating this operation for various
values of $w$ as well as of the free parameters of the model, one
finds the required equation of state, albeit only
numerically~\cite{neutral,enon0}.

For this model, it was shown that, in the electric regime where the
current is timelike, the current diverges logarithmically when the
state parameter approaches the current--carrier mass squared, $m_*^2$
say. Using this property, it was then possible to propose a best fit
to the otherwise numerical equation of state~\cite{Carter-Peter}, fit
which is amazingly good for almost all values of the state
parameter. In particular, including a divergence in the electric
regime was shown to also imply a current saturation in the magnetic
regime. In the Lagrangian formalism, it reads, setting the string's
characteristic mass scale to $m$,
\begin{equation} 
{\cal L}\{ w\} = -m^2 -{m_*^2\over 2} \ln \left\{ 1 + {w\over m_*^2}
\right\} ,\label{fitL}\end{equation}
which, upon using Eqs.~(\ref{wK}) and (\ref{Lamb-}), 
provides ${\cal K}$ as a function of
$\chi$ in the form
\begin{equation} {\cal K} = 2 {d\Lambda\over d\chi} = 1+{w\over m_*^2}=
{1-\sqrt{1 - 4 \chi /m_*^2} \over 2\chi
/m_*^2},\label{Kold}\end{equation}
where, in the last equality, use has been made of Eq.~(\ref{wK}) and 
the minus sign in front of the squared 
root ensures that
${\cal K} \to 1$ 
when $\chi \to 0$ (the Goto--Nambu limit of no current).
Integrating Eq.~(\ref{Kold}), and normalizing in such a way that
$\Lambda \{ 0\} = {\cal L} \{ 0\} =-m^2$, yields $\Lambda$ as
\begin{eqnarray} \Lambda = -m^2 +{m_*^2\over 2} & & \left[ 1-\sqrt{1-4 \chi
/m_*^2} \right. \nonumber \\ & & + {1\over 2} \left. \ln \left(
{\chi\over m_*^2} {1+\sqrt{1-4 \chi /m_*^2}
\over 1-\sqrt{1-4 \chi /m_*^2}} \right)\right]
,\label{Lambdold} \end{eqnarray} 
in which it now suffices to
incorporate the shift~(\ref{start}) to account for the inclusion of
electromagnetic correction at first order 
[and note that now $\cal K$ is modified to ${\cal K} + 
\lambda  q^2$, as one clearly sees from Eqs.~(\ref{calk-})
and (\ref{KK})].  As should be clear on this particular example,
getting back to the original Lagrangian formalism would be a very
awkward task, and in fact does not lead to any analytically known form
for ${\cal L}\{ w\}$. This unpleasant feature however is not much
bother since the dual formalism is avalaible.

An important point needs be noted at this stage: it concerns the
relevant dimensionless parameters. The model~(\ref{fitL}) in fact
depends solely on one such parameter, namely the ratio
\begin{equation} \alpha = \left( {m\over m_*} \right) ^2,\label{alpha}
\end{equation}
which, in a reasonnable cosmic string microscopic model~\cite{witten},
would be at least of order unity, and in many
applications~\cite{vortons} largely in excess of unity. As it was
shown~\cite{US} in a previous study not including electromagnetic
corrections that the results for the vortons themselves were not in
any essential way dependent on $\alpha$ as long as $\alpha\agt 1$, we
shall consider the case $\alpha =1$ on the figures that follow.

As an illustration, Eqs.~(\ref{wK}) and (\ref{Lamb-}) have been used
to calculate the equation of state $U(\nu )$ and $T(\nu )$ for various
values of the electromagnetic correction parameter $\lambda q^2$, and
they are exhibited on Fig.~\ref{eosfig}. Similar figures can be found
in Ref.~\cite{neutral,enon0}, with the same axis $\nu$ (scales are
different because not normalized in the same way) for the numerically
computed equation of state in the Witten bosonic superconducting
cosmic string field--theoretic model. 
On this figure, $U$ and $T$ are plotted as
functions of $\nu$, which is defined as the square root of the state
parameter $w$: $\nu = $~Sign$(w) \sqrt{|w|}$. Its meaning is very
simple: for a straight string lying along the $z$ axis say, one can
set the phase of the current carrier as $\varphi = k z -
\omega t$, and there exist a frame in which $\nu$ is either $k$ or
$\omega$, i.e. it represents the momentum of the current--carrier along
the string's direction, or its energy. The electromagnetic correction
in this case is seen to enlarge the picture: a small (or vanishing)
correction yields the usual form of the equation of state where the
tension (hence $c_{_E}^2$) goes to zero for large negative $w$ (phase
frequency threshold)~\cite{neutral}, 
and $c_{_L}^2$ vanishes on the magnetic side for
$w=w_s$ (saturation).  The threshold becomes more and more negative
with increasing $\lambda q^2$, and the saturation point is reached for
larger values of $w$; both these remarks show that inclusion of
electromagnetic corrections can be interpreted as a rescaling of
$w$, which in Fig.~\ref{eosfig} is equivalent to rescaling the $x$
axis. 

\section{Circular motion in flat space}

We now restrict our attention to the motion of a circular vortex ring
in flat space. The analysis in this case has already been
done~\cite{US}, so we only need to summarize the results, and
eventually rephrase them in terms of $\Lambda$ instead of ${\cal L}$.

\subsection{Equations of motion}

The background and the solution admit two Killing vectors, one
timelike $k^\mu$ normalized through
\begin{equation} k^\mu {\partial \over \partial x^\mu} = {\partial
\over \partial t},\end{equation}
$t$ being a timelike coordinate, and one spacelike $\ell^\mu$
\begin{equation} \ell^\mu {\partial\over\partial x^\mu} = 2\pi
{\partial\over\partial\phi},\end{equation} with $\phi\in [0,2\pi]$ an
angular coordinate; both $t$ and $\phi$ are ignorable.The length
$\ell$ of the string loop is then given by
\begin{equation} \ell^2 = \ell^\mu\ell_\mu,\end{equation}
while its total mass $M$ and angular momentum $J$ are defined by
\begin{equation} M\equiv  -\oint dx^\mu \epsilon_{\mu\nu}
{\overline T^\nu}_\rho k^\rho ,\label{M}\end{equation}
and
\begin{equation} J\equiv  (2\pi)^{-1} \oint dx^\mu \epsilon_{\mu\nu}
{\overline T^\nu}_\rho \ell^\rho .\label{J}\end{equation}
So long as we do not consider radiation of any kind (a requirement
equivalent with the demand that $k^\mu$ and $\ell^\mu$ be Killing
vectors also for the string configuration), these are conserved. Note
also that the relation $J=NZ$ holds.

\begin{figure}
\centering
\epsfig{figure=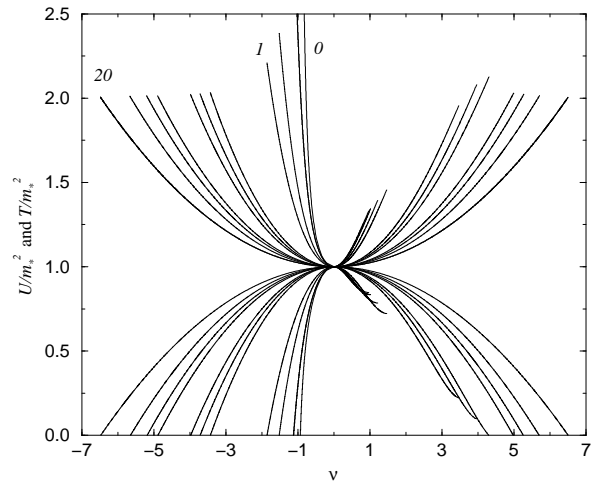, width=9cm}
\caption{Variation of the equation of state with the electromagnetic
self--correction $\lambda q^2$. It relates the energy per unit length
$U$ (upper set of curves) and the tension $T$ (lower set of curves), 
both in units of $m_*^2$, the current--carrier
mass, and is plotted against $\nu$, which is the square root of the
state parameter $w$. Values used for this correction are in the set
$[0, 0.1,0.5, 1, 2, 5, 7, 8, 9, 10, 20]$, and the figure is calculated
for $\alpha=1$. Increasing the value of $\lambda q^2$ enlarges the
corresponding curve in such a way that for very large values (in this
particular example, it is for for $\lambda q^2 \geq 7$), the tension on
the magnetic side becomes negative before saturation is reached.
\label{eosfig}}
\end{figure}

Other quantities need be introduced, related with the integer numbers
$N$ and $Z$, in which it turns out to be convenient to include the
scale parameter $\kappa_{_0}$ (or equivalently $\tilde\kappa_{_0}$) in
the following clearly dual definitions
\begin{equation} B\equiv N/\sqrt{-\tilde\kappa_{_0}} \ \ \ , \ \ \ 
C \equiv Z/\sqrt{\kappa_{_0}}.\end{equation}

Now specifying to the particular case of a flat spacetime background
in which the circular string is confined on a plane so that we can
use circular coordinates $\{ r, \theta, \phi ,t \}$ 
and set $\theta =\pi
/2$, $t=M\tau$ and $\phi = \sigma$ (recall $\tau$ and $\sigma$ are the
respectively timelike and spacelike internal coordinates), 
use of equations (\ref{win}) imply that the phases vary like
\begin{equation} \psi = s(t)+Z\phi \ \ \ ,\ \ \ \varphi = f(t) + N\phi,
\end{equation}
with $s$ and $f$ functions only of time and expressible 
in terms of the conserved numbers $B$ and $C$ through
\begin{equation} \dot s = {B\over {\cal K}} {\sqrt{1-\dot r^2}\over 2\pi r
\sqrt{-\tilde\kappa_{_0}}} \ \ \ , \ \ \ \dot f = C {\cal K}
{\sqrt{1-\dot r^2}\over 2\pi r \sqrt{\kappa_{_0}}},\end{equation} a dot
meaning a derivative with respect to the time coordinate $t$. The
variation of the string's radius $r=\ell /2\pi$ follows from the
equation~\cite{US}
\begin{equation} M\sqrt{1-\dot r^2} = \Upsilon (r)
,\label{eomr}\end{equation} 
from which we conclude that the string
evolves in a self--potential $\Upsilon (r)$. Thus, it
suffices to know the form of this potential to understand completely
the proto--vorton dynamics. It is given, in terms of our variables,
by~\cite{US}
\begin{equation} \Upsilon = {B^2 \over {\cal K} \ell} - \Lambda \ell ,
\label{upsilon} \end{equation}
while the string's circumference reads
\begin{equation} \ell^2 = {1\over \chi} \left( {B^2 \over {\cal K}^2 }
- C^2\right).\label{ell}\end{equation}

In order to express the results, it is simpler to rescale everything
by means of $\chi_r = \chi /m_*^2$, $\ell_r = m_* \ell / |C|$,
$\Lambda _r = \Lambda /m_*^2$, ${\cal L} _r = {\cal L} / m_*^2$, and
$\Upsilon_r = \Upsilon /(m_* |C|)$ so that all quantities of interest
are dimensionless and depend only on three arbitrary also
dimensionless parameters, namely $\alpha$, which we discussed already,
$b$, defined by $b^2 = B^2/C^2$ and through which one expresses the
timelike or spacelike character of the current 
[from Eq.~(\ref{ell})], and the most important
parameter here, namely $\lambda q^2$. We are now ready to examine the
actual electromagnetic correction to a string loop dynamics 
at first order in the coupling $j_\mu A^\mu$.

\subsection{The self potential}

The potential $\Upsilon_r$ as a function of $\ell_r$ is derivable by means
of first expressing $\Upsilon_r$ and $\ell_r$ as functions of the state
parameter $\chi_r$ through Eqs.~(\ref{upsilon}) and (\ref{ell}). In
order to do this, one needs to know the range in which $\chi_r$
varies, range given by the requirements~(\ref{stabc-}), which can be
rephrased into the following constraints:
\begin{eqnarray}
2\alpha-\lambda q^2 \chi_r -1+\sqrt{1-{4 \chi_r}} 
\nonumber\\  %%
-\frac{1}{2} \ln\left\{ \chi_r {1+\sqrt{1-{4 \chi_r}}\over
1-\sqrt{1-{4 \chi_r}} }\right\}
&>& 0
\hskip 0.1 cm
{\rm [ {\it T} > 0 \, \, magnetic ],} 
\label{constraint1} \\
2\alpha+\lambda q^2 \chi_r 
\; \; \; \; \; \; \; \; \; \; \; \; \; \; \; \; \; \; \; \; \; \;
\; \;
\phantom{.}
\nonumber\\  %%
-\frac{1}{2} \ln\left\{ \chi_r {1+\sqrt{1-{4 \chi_r}}\over
1-\sqrt{1-{4 \chi_r}} }\right\}
&>& 0
\hskip 0.1 cm
{\rm [ {\it T} > 0 \, \, electric ],} 
\label{constraint2} 
\end{eqnarray}
\begin{equation}
{ 1-\sqrt{1-{4 \chi_r}}+\lambda q^2 2 \chi_r \sqrt{1-{4 \chi_r}}
\over
\sqrt{1-{4 \chi_r}}\left[ 1-\sqrt{1-{4 \chi_r}}+\lambda q^2 2
\chi_r  \right] }
> 0
\hskip 0.2 cm
[ c_{_{\rm L}}^{\,2} > 0 ].
\label{constraint3} 
\end{equation}

\begin{figure}
\centering
\epsfig{figure=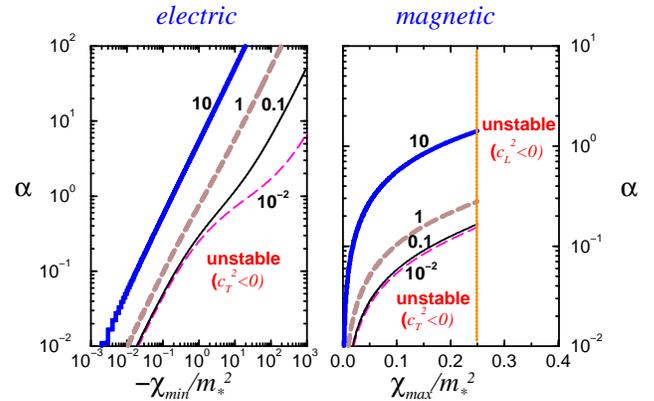, width=9cm}
\vspace{-2cm}
\caption{Constraints yielding minimum and maximum values for 
the normalized current $\chi / m_\ast^2$, according to 
Eqs.~(\ref{constraint1}), (\ref{constraint2}), and
(\ref{constraint3}), for values $\lambda q^2 = 10^{-2}$ to $10$.
In the figure on the left, for each particular value of $\lambda q^2$,
the unstable region (where $c_{_{\rm T}}^{\,2} < 0$) 
lies below the corresponding curve. 
The same is true for the figure on the right, but in addition 
$\chi / m_\ast^2 < 1/4$ for otherwise $c_{_{\rm L}}^{\,2} < 0$.
\label{constraintsfig}}
\end{figure}

The first and second of these constraints specify the condition that
the extrinsic `wiggle' squared perturbation velocity must be positive
(and therefore also the tension $T > 0$) in both the magnetic and the
electric regimes (i.e., for spacelike and timelike currents,
respectively).  The third constraint Eq.~(\ref{constraint3}) demands
that the longitudinal `woggle' squared perturbation velocity be
positive and is always satisfied provided $\chi_r < 1/4$.  We plot
this as the vertical line in the picture on the right of
Fig.~\ref{constraintsfig}, the allowed range of $\chi_r$ values being to
the left of it. These constraints were used in the calculation of
Fig.~\ref{eosfig}.

The first thing to determine in order to plot $\Upsilon_r (\ell_r )$ is
whether the current is timelike or spacelike. This is achieved by
looking at Eq.~(\ref{ell}) which states that the sign of $\chi$, and
hence the timelike or spacelike character of the current, is also the
sign of $b^2-{\cal K}^2$. Now Eq.~(\ref{Kold}), modified to account
for (\ref{start}), shows that the range of variation of ${\cal K}$
is
\begin{equation} \chi \geq 0 \ \ \hbox{(magnetic)} \Longrightarrow
\ 1+\lambda q^2 \leq {\cal K} \leq 2+\lambda q^2 ,
\label{mag-range}\end{equation}
and
\begin{equation} \chi \leq 0 \ \ \hbox{(electric)} \Longrightarrow
\ \lambda q^2 \leq {\cal K} \leq 1+\lambda q^2 .
\label{elec-range}\end{equation}
Therefore, it is only possible for $\chi$ to be positive if $b\geq 1+
\lambda q^2$, and negative otherwise. So we deal with a magnetic
configuration or an electric configuration depending on whether $b$ is
respectively greater or less than its critical value $b_c = 1+\lambda
q^2$. Since $\chi$ cannot change sign dynamically, by fixing $b$ one
also fixes the character of the current and, where the physical
constraints (\ref{constraint1}) to (\ref{constraint3}) cease to be
satisfied, the curves plotted in the various figures end.  As the
quantity $b_c$ was unity in the decoupled case, we see that
electromagnetic corrections can modify the nature of the current for a
given set of integer numbers $Z$ and $N$.

\begin{figure}
\centering
\epsfig{figure=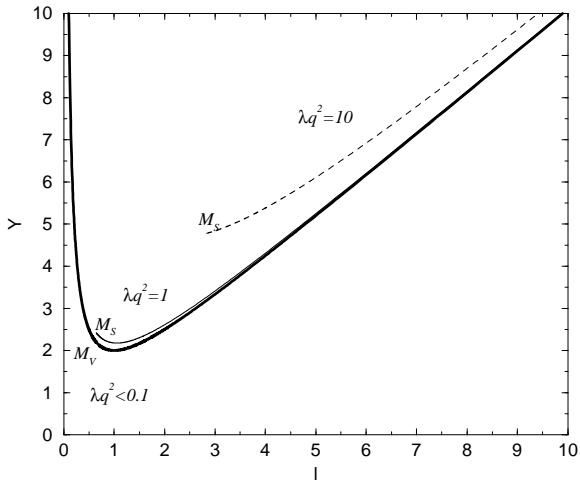, width=9cm}
\caption{Variations of the self potential $\Upsilon_r$ with the ring's
circumference $\ell_r$ and the electromagnetic self coupling $\lambda
q^2$ for $\alpha = b =1$. The thick curve stands for various values of
$\lambda q^2 < 0.1$ for which they are undistinguishable, and in the
``safe'' zone; the minimum value of $\Upsilon_r$ is then $M_v$, the
vorton mass. $\Upsilon_r$ for the same parameters $\alpha$ and $b$,
this time for $\lambda q^2 =1$ is represented as the full thin line,
where it is clear that we now are in a ``dangerous'' zone where the
potential has a minimum (new value for $M_v$) but now terminates at
some point where it equals $M_s$. Finally the dashed curve represents
the potential for $\lambda q^2 = 10$, an unrealistically large value,
and this time the curve terminates even before reaching a minimum:
this is a ``fatal'' situation for all loops with such parameters will
eventually decay.\label{Yfig}}
\end{figure}

Once the nature of the current is fixed, it is a simple matter to
evaluate the potential $\Upsilon_r$, and it is found, as in
Ref.~\cite{US}, that three cases are possible, depending on the values
of the free parameters, namely the so--called ``safe'', ``dangerous''
and ``fatal'' cases. They correspond to whether the thin string
description holds for all values of the allowed parameters or not, as
illustrated on Fig.~\ref{Yfig}.

The general form of the potential $\Upsilon_r (\ell_r )$ exhibits a
minimum and two divergences, one at the origin which prevents a
collapse of the loop, and one for $\ell_r\to\infty$ which holds the
loop together and is responsible for the confinement
effect~\cite{US}. The latter divergence, going like
$\Upsilon_r\sim\ell_r$, occurs whatever the underlying parameters may
be and is mainly due to the fact that it requires an infinite amount
of energy to enlarge the loop to infinite size, its energy per unit
length being bound from below ($U\geq m^2$). The divergence at $\ell_r
=0$ is however not generic, as is shown on Fig.~\ref{Yfig}, since for
some sets of parameters, the quantity $\ell_r$ does not take values in
its entire range of potential variations $[0,\infty [$. This can be
seen as follows.

\begin{figure}
\centering
\epsfig{figure=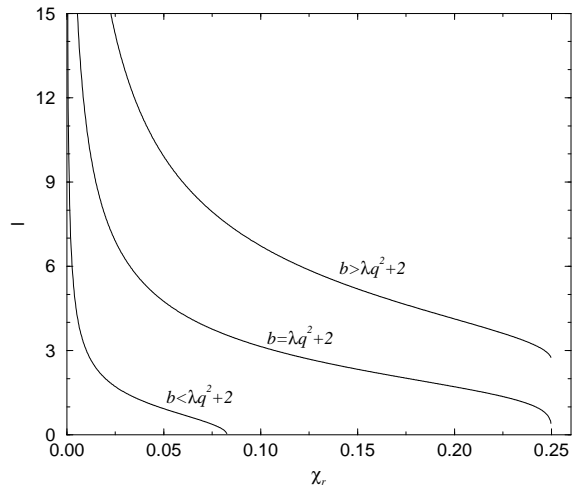, width=9cm}
\caption{The characteristic behavior of $\ell_r$ as a function of
$\chi_r$ for $\lambda q^2=1$ in the magnetic regime where $\chi_r >0$:
the curve down, indicated $b<\lambda q^2+2$ is for $b=2.1$, and the
upper curve for $b=5$. They all diverge around $\chi_r
=0$.\label{l2posfig}}
\end{figure}

On the magnetic side, the function $\ell_r ^2 (\chi _r)$ ranges from
$+\infty$ in the limit where $\chi _r\to 0^+$, to $4[b^2 - (\lambda
q^2 +2)^2 ]/(\lambda q^2 +2)$ for $\chi _r\to 1/4$ as sketched on
Fig.~\ref{l2posfig}. Depending therefore on whether $b$ is less or
greater than $(\lambda q^2 +2)$, the $\ell_r =0$ limit will or will
not exist. In the former case, the potential $\Upsilon_r$, which
diverges for $\ell_r \to 0$, will have the form indicated as the thick
curve on Fig.~\ref{Yfig}, and the string loop solution is in a
``safe'' zone.  If $b>\lambda q^2 + 2$, then the minimum value for
$\ell_r ^2$ is non zero so the potential terminates at some point,
which can be either sufficiently close to the origin that the minimum
for $\Upsilon_r$ can be reached (``dangerous'' zone) or not (``fatal''
zone). In the former case, the resulting configuration may reach an
equilibrium state of mass $M_v$ (when radiation is taken into account,
such a configuration will eventually loose enough energy to settle
down into a vorton state) provided its mass $M$ is less than that
obtained for $\chi_r = 1/4$, $M_s$ say, with the same set of
parameters, whereas it will enter a regime in which the thin string
description is no longer valid if $M > M_s$. Finally, there is also
the possibility that no minimum of $\Upsilon_r (\ell_r )$ is
attainable on the entire available range for $\ell_r$; this is called
``fatal'' because, whatever the value of $M$, the loop again ends up
in the region where no string description holds anymore. When this
happens, the topological stability of the vortex can be removed
dynamically and the quantum effects make the loop decay into a burst
of Higgs particles.

\begin{figure}
\centering
\epsfig{figure=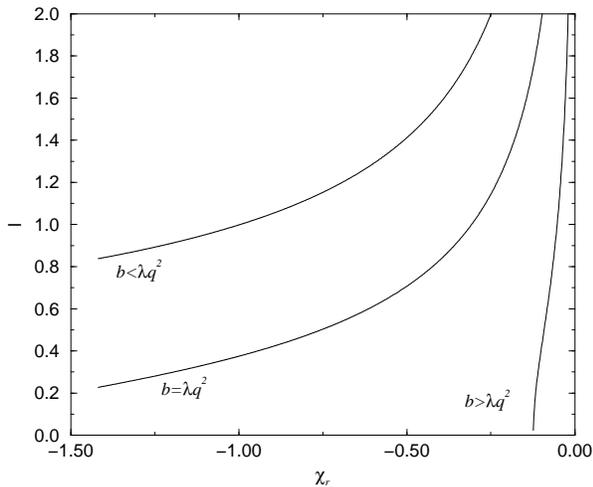, width=9cm}
\caption{The characteristic behavior of $\ell_r$ as a function of
$\chi_r$ for $\lambda q^2 =1$ in the electric regime where $\chi_r
<0$: The curve down is for a ``safe'' configuration with $b=1.9 >
\lambda q^2$, whereas the upper curve is ``fatal'', with $b=0.1 < \lambda
q^2$. The two upper curves terminate at the point where the string
tension becomes negative and the string is unstable with respect to
transverse perturbations.\label{l2negfig}}
\end{figure}

The electric regime presents roughly the same features of having
``safe'', ``dangerous'' and ``fatal'' zones, although for slightly 
different reasons: 
the magnetic case ends either when $c_{_{\rm L}}^{\,2} \to 0$
or $c_{_{\rm T}}^{\,2} \to 0$ whereas the electric case does so only 
in the case $c_{_{\rm T}}^{\,2} \to 0$.
On Fig.~\ref{l2negfig} is sketched the function $\ell_r (\chi
_r)$ for $\chi _r <0$ and various values of the parameters $b$, given
$\lambda q^2$. What happens in the electric regime is that the
limiting case this time is for $b=\lambda q^2$, $\ell_r ^2$ behaving as
$(b^2 - \lambda^2 q^4)/(\lambda^2 q^4 \chi_r)$ as $\chi_r\to -\infty$;
thus, if $b<\lambda q^2$, as $\chi _r$ is negative, $\ell_r^2$ is always
non zero and there must exist a value for $\chi_r$ such that
the string's tension vanishes, and the loop itself becomes unstable
with respect to transverse perturbations. Such a loop would clearly
not form a vorton. On the other hand, if $b>\lambda q^2$, then
$\ell_r^2$ can get to zero, and for some values of the
parameters (unfortunately the range is not derivable analytically), it
will do so before the tension vanishes. The corresponding loop might
then end as a vorton.

Finally the safe zone, for all regimes taken together, is limited to
\begin{equation} \lambda q ^2 \leq b \leq \lambda q^2 + 2 ,
\label{limits}\end{equation}
a condition which is increasingly restrictive as $\lambda q^2$
increases, and may even forbid vorton formation altogether for a very
large coupling.

\section{Conclusions}

We have exhibited explicitely the influence of electromagnetic self
corrections on the dynamics of a circular vortex line endowed with a
current at first order in the coupling between the current and the
self--generated electromagnetic field, i.e., neglecting radiation.
This is necessary before any radiation can be taken into account and
evaluated, a task which is still to be done. Moreover, use of the
duality formalism developed by Carter~\cite{Formalism} has been made
and shown to be especially useful in this particular case in the sense
that it enabled us to derive the dynamical properties of a
proto--vorton configuration analytically. It is to be expected that
such a dual calculation will prove indispensable when evaluation of
the higher electromagnetic orders will be performed.

\begin{figure} 
\centering 
\epsfig{figure=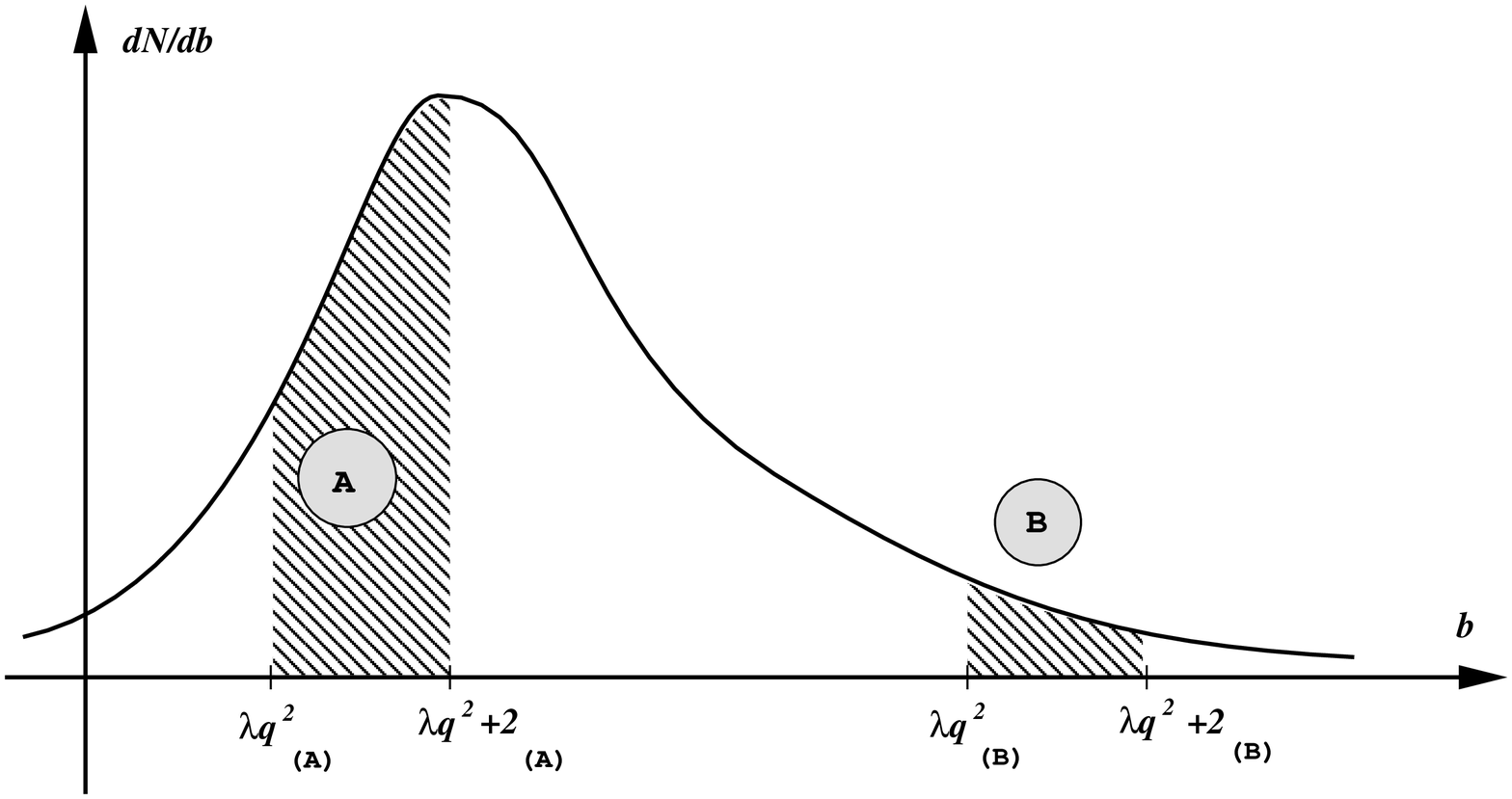, width=9cm} 
\caption{A possible way out of the vorton excess problem: a sketch of a
distribution of loops with $b$, and ``safe'' intervals
[cf. Eq.~(\ref{limits})] for different values of $\lambda q^2$.  It is
clear that the actual number density of ensuing vortons, at most
proportional to the shaded areas, will depend quite strongly on the
location of the safety interval. Note also that this electromagnetic
correction may reduce drastically the available phase space for vorton
formation since the maximum of the $dN/db$ distribution is usually
assumed to be peaked around $b=1$.
\label{fig-sketch}} 
\end{figure} 

We have shown that most of the conclusions of our previous paper on
that subject actually hold when electromagnetism is accounted for, at
least at this order, with the result, perhaps not intuitively obvious
from the outset, that this self--interaction tends to destabilise the
string loops towards states for which a classical string description
does not hold, configurations which are expected to decay into the
string constituents (the Higgs field in particular) when quantum
effects are taken into account. 

As is clear from Eq.~(\ref{limits}), increasing the electromagnetic
correction is equivalent to reducing the available phase space for
vorton formation, as $b$ of order unity is the most natural
value~\cite{vortons}, situation that we sketch in
Fig.~\ref{fig-sketch}. On this figure, we have assumed a sharply
peaked $dN/db$ distribution centered around $b=1$; with $\lambda q^2
=0$, the available range for vorton formation lies precisely where the
distribution is maximal, whereas for any other value, it is displaced
to the right of the distribution. Assuming a gaussian distribution,
this effect could easily lead to a reduction of a few orders of
magnitude in the resulting vorton density, the latter being
proportional to the area below the distribution curve in the allowed
interval.  This means that as the string loops contract and loose
energy in the process, they keep their ``quantum numbers'' $Z$ and $N$
constant, and some sets of such constants which, had they been
decoupled from electromagnetism, would have ended up to equilibrium
vorton configurations, instead decay into many Higgs particles,
themselves unstable. This may reduce the cosmological vorton excess
problem~\cite{vortons} if those are electromagnetically charged.

The present analysis, because of its being restricted to exactly
circular configurations, is not sufficient to provide general
conclusions as to whether vortons will form or not for whatever
original loop shapes, but clearly indicates that even though the
cosmological vorton problem~\cite{vortons} cannot be solved by means
of this destabilizing effect, it may well have been slightly
overestimated.

\section*{Acknowledgments}

We would like to thank B.~Carter for his insights on the
matters delt with here, and M.~Sakellariadou for many stimulating
discussions.
The research of A.G. was supported by the
Fondation Robert Schuman.  He also acknowledges partial financial
support from the programme Antorchas/British Council (project
N$^\circ$ 13422/1--0004).

\end{document}